\documentclass[preprint,aps]{revtex4}
\usepackage[dvips]{graphicx}
\usepackage{graphicx}
\usepackage{amsfonts}
\usepackage{bm}
\usepackage{amsmath}
\usepackage{amssymb}
\usepackage{color}
\usepackage[all]{xy}

\def\be{\begin{equation}}
\def\ee{\end{equation}}
\def\bea{\begin{eqnarray}}
\def\eea{\end{eqnarray}}

\begin{document}
\title{Stochastic Einstein equations with fluctuating volume}
\author{Vladimir Dzhunushaliev$^{1,2}$ and Hernando Quevedo$^{3,4}$}
\email{v.dzhunushaliev@gmail.com, quevedo@nucleares.unam.mx}

\affiliation{
$^1$ Physical-Technical Faculty, al -- Farabi Kazakh National University, al -- Farabi av. 71, 050040, Almaty, Kazakhstan\\
$^2$ IETP, al -- Farabi Kazakh National University, 050040, Almaty, Kazakhstan \\ 
$^3$ Instituto de Ciencias Nucleares,
Universidad Nacional Aut\'onoma de M\'exico,
 AP 70543, M\'exico, DF 04510, Mexico\\
$^4$ Dipartimento di Fisica and ICRA, Universit\`a di Roma "La Sapienza",  Piazzale Aldo Moro 5, I-00185 Roma, Italy\\
}

\begin{abstract}
We develop a simple  model to study classical fields on the background of a fluctuating spacetime volume. It is applied 
to formulate the stochastic Einstein equations with a perfect-fluid source. We investigate the particular case of a stochastic
Friedmann-Lema\^itre-Robertson-Walker cosmology, and show that the resulting field equations can lead to solutions which avoid the 
initial big bang singularity. By interpreting the fluctuations as the result of the presence of a quantum spacetime, we conclude 
that classical singularities can be avoided even within a stochastic model that include quantum effects in a very simple manner.

\end{abstract}

\maketitle

\section{Introduction}

The idea of taking into account stochastic effects in Einstein gravity in order to investigate quantum gravitational effects is not new. In fact, several approaches have been proposed. For instance in \cite{moff97}, a stochastic theory of gravity was proposed in which the gravitational and matter fields are treated as probabilistic quantities. 

A different approach is the so-called semiclassical stochastic gravity  \cite{hv03,hv04}. 
In a full theory of quantum gravity, one would expect that a metric operator $\hat g_{\mu\nu}(x)$ is related to the energy-momentum operator $\hat T_{\mu\nu}(x)$ in a 
(still unknown) consistent manner. A simplified theory consists in 
considering the classical metric $g_{\mu\nu}(x)$, instead of the metric operator, as a classical variable, and the expectation value of the energy-momentum operator 
$<\hat T _{\mu\nu}(x)>$, instead of the operator itself. This semiclassical theory of gravity has been used successfully in black hole physics and in cosmological scenarios.  As long as the gravitational field is assumed to be described by a classical metric, this semiclassical theory is the only dynamical equation for this
metric. Indeed, the metric couples to the the matter fields through the energy-momentum tensor and for a given quantum state the expectation value is the only physical observable that one can construct. However, since no full theory of quantum gravity is known, it is not clear how to determine the limits of applicability of the semiclassical theory. It is expected that this theory should break down at Planck scales at which the quantum effects of gravity should not be ignored. Also, the 
semiclassical theory is expected to break down when the quantum fluctuations of the energy-momentum tensor become large and non-negligible. 

Stochastic semiclassical theory is an attempt to take into account in a consistent manner the quantum fluctuations of the energy-momentum tensor \cite{hv03,hv04}.  Formally, to consider quantum fluctuations at the level of the field equations one adds a stochastic function $\xi(x)$ to the expectation value of the energy-momentum tensor so that the classical metric $g_{\mu\nu}(x)$ couples now to the sum   $<\hat T _{\mu\nu}(x)>+ \xi(x)$. The resulting semiclassical Einstein-Langevin  equations determine the dynamic evolution of the classical metric which acquires now a stochastic character. It is not easy to analyze physical situations in semiclassical stochastic gravity because it is necessary to handle not only the conceptual and technical problems of describing quantum fields in curved spacetimes, but also the stochastic function source that accounts for the quantum fluctuations. This is probably one of the reasons why research in this direction \cite{srh03,rv08,slb13}  is less intensive than in  other formal quantization methods of gravity \cite{carlip}. 

Nevertheless, a concrete set of stochastic Einstein equations was formulated recently in 
\cite{dzu11}. In this approach, the dynamical variables correspond to  stochastic quantities that 
fluctuate according to some probability density.  To obtain the probability density and the field equations, it is assumed that the Ricci flow is a statistical 
system and every metric in the Ricci flow corresponds to a microscopical state.

In this work, we propose an alternative model to take into account quantum fluctuations. 
All the approaches used to quantize gravity try to show or assume that at some scales spacetime becomes quantized, i.e., it can have only discrete values. 
This expectation has been corroborated in loop quantum gravity where area and volume quantum operators have been obtained with discrete spectra. Although still many technical and conceptual issues remain to be solved \cite{rov08}, the quantization of the spacetime volume seems to be well established. This implies that quantum fluctuations should appear that would affect the geometric structure of spacetime. We propose a  model in which the stochastic character of spacetime is taken into account in a very simple manner at the level of the field equations. In fact, we will see that it is possible to introduce a stochastic factor into the field equations so that the structure of spacetime becomes affected by the presence of fluctuations.  Thus, our model consists of a classical theory on the background of a quantum fluctuating volume. We will apply our model to the Friedmann-Lema\^itre-Robertson-Walker (FLRW) cosmology in order to explore the question about the possibility of avoiding the initial big bang singularity. We will find the conditions under which the avoidance is possible.

This work is organized as follows. In Sec. \ref{sec:mod}, we derive the stochastic classical field equations in the case of a perfect-fluid source, and study its properties. We show that the field equations can be interpreted as containing an effective cosmological constant with dynamic properties which arise as a 
consequence of quantum fluctuations. Interestingly,  stiff matter turns out to remain unaffected by the presence of quantum fluctuations. In Sec. \ref{sec:flrw}, we investigate the stochastic classical equations in the case of a FLRW cosmological model, and show that the big bang singularity can be avoided in a large class of cosmological models. Finally, in Sec. 
\ref{sec:con}, we discuss our results.


\section{Stochastic field equations for a perfect fluid}
\label{sec:mod}

One of the goals of quantizing gravity is to show that there exists a limit at which our classical understanding of spacetime breaks down, and it must be 
replaced by a different concept, say, a quantum spacetime. This is the case of loop quantum gravity in which area and volume operators arise which are characterized by discrete spectra. This is a clear indication of the presence of a quantum spacetime. Other quantization methods even start from the assumption that at some level 
spacetime is quantum, and different approaches can be used to implement this idea into a mathematical model \cite{carlip}. 

On the other hand, it has been well established in physics that quantum quantities fluctuate. It follows that spacetime fluctuations should be a consequence of quantization of gravity. In fact, as mentioned above, the implementation of this idea led to the formulation of stochastic semiclassical gravity. The resulting 
Einstein-Langevin equations, however, are not easy to be handled and many technical difficulties appear because in the corresponding field equations 
it is necessary to consider the expectation value of the energy-momentum tensor and the stochastic functions, representing the spacetime fluctuations. In this work, 
we propose a simple model in which the stochastic nature of spacetime is modeled by a stochastic factor at the level of the classical field equations. One could 
say that we will propose a simple model of stochastic classical gravity which allows us to investigate in an effective manner 
the consequences of a fluctuating quantum spacetime. 

To be more specific, let us consider the general action of Einstein's gravity with a matter Lagrangian $\mathcal L _m$,
\begin{equation}\label{1-10}
  \mathcal L = \sqrt{-g} \left(
    - \frac{1}{2 \varkappa} R + \mathcal L_m 
  \right) \ .
\end{equation}
The corresponding field equations are usually obtained by imposing the variational principle in the form
\be
\delta \mathcal L =0= \sqrt{-g} \left(  \frac{1}{2 \varkappa} \delta R + \delta\mathcal L_m \right) 
+    \left( - \frac{1}{2 \varkappa} R + \mathcal L_m  \right) \delta \sqrt{-g} \ ,
\label{feq}
\ee
where the variation of the determinant of the metric is given by
\be
 \delta \sqrt{-g} = - \frac{1}{2} g_{\mu\nu} \delta g^{\mu\nu}  .
\label{vdet}
\ee
Now, we want to consider the contribution of a fluctuating spacetime to the field equations (\ref{feq}). 
Clearly, in a fluctuating spacetime, the spacetime volume $\sqrt{-g}d^4x$ will fluctuate and the variation $\delta \sqrt{-g} d^4x$ should take this 
into account. Then, the simplest possible way to consider fluctuations in the variation of the spacetime volume is by introducing a stochastic factor 
$  \mathfrak{s} $ into the variation (\ref{vdet}), i.e., 
\be
 \delta \sqrt{-g} = - \frac{1}{2}   \mathfrak{s}  g_{\mu\nu} \delta g^{\mu\nu} \ .
\label{vdets}
\ee
At the level of the field equations this is equivalent to replacing   $g_{\mu\nu}$ by $ \mathfrak{s}  g_{\mu\nu}$. Consequently, in the case of a 
perfect-fluid source with pressure $p$ and energy density $\epsilon$, we obtain 
\be
R_{\mu\nu} - \frac{1}{2}  \mathfrak{s} R g_{\mu\nu} = - 8 \pi [ (p+\epsilon) u_\mu u_\nu - \mathfrak{s}\, p\, g_{\mu\nu} ]\ .
\label{see}
\ee
Thus, we see that within the context of this simple model a fluctuating spacetime  leads to a non-trivial modification of the field equations. 
The question is whether the presence of the stochastic factor in the field equations leads to physical consequences that could be associated with the quantum nature of spacetime. We will analyze this problem in the next section.  

We will now analyze the general properties of the stochastic classical Einstein equations (\ref{see}). First, we notice that in the vacuum case with arbitrary 
$ \mathfrak{s}\neq \frac{1}{2}$, the field equations reduce to $R_{\mu\nu}=0$. This implies that any possible stochastic effects might be detected only in the presence of matter.
This is certainly a consequence of the simplicity of the model we are considering in this work; the spacetime fluctuations affect only the matter and not the spacetime itself. 

The stochastic field equations (\ref{see}) can be rewritten as 
\be
G^{\mu\nu} = -8\pi [ (p+\epsilon) u^\mu u^\nu - \mathfrak{s} p g^{\mu\nu} ] + \frac{1}{2}( \mathfrak{s}-1) R g^{\mu\nu} \ ,
\label{eins}
\ee
where  $G^{\mu\nu}$ is the Einstein tensor. From the Bianchi identities we know that the Einstein
tensor is divergence free, i.e.,  $G^{\mu\nu}_{\ ;\nu} = 0$, which usually leads to the 
conservation law $T^{\mu\nu}_{\ ;\nu} = 0$.  However, in the above case the presence of the stochastic
factor $ \mathfrak{s}$ implies a violation of the conservation law. Indeed, the Bianchi identities imply in this case that
\be
-8 \pi [ (p+\epsilon) u^\mu u^\nu - p g^{\mu\nu} ]_{;\nu} =\frac{1}{2}(1- \mathfrak{s})\left (R  + 16 \pi p\right)_{,\nu} g^{\mu\nu}\ ,
\label{bianchi}
\ee
so that the term on the right-hand side of the above equation is responsible for the violation of the conservation law. The conservation law 
is recovered not only in the trivial case $\mathfrak{s}=1$, but also for 
\be
R= - 16 \pi p \ ,
\label{stiff}
\ee
which corresponds to stiff matter. Indeed, 
calculating the trace of the field equations, we obtain the relationship
\be
(1-2\mathfrak{s}) R = -8\pi [ \epsilon +(1-4 \mathfrak{s}) p]\ ,
\label{spur}
\ee
which, when replaced in Eq.(\ref{stiff}), leads to the equation of state of stiff matter, $\epsilon=p$. Furthermore, notice 
that in the particular case $ \mathfrak{s}=1/2$, from the trace equation (\ref{spur})  the equation of state for stiff matter arises, independently of the value of $R$. 
This implies that stiff matter is not affected by the presence of the stochastic factor in the metric. 

Finally, let us mention that the stochastic field equations (\ref{see})  can also be written as 
\be
G^{\mu\nu} = -8\pi [ (p+\epsilon) u^\mu u^\nu - p g^{\mu\nu} ] +\frac{1}{2} ( \mathfrak{s}-1) \left(R + 16 \pi p \right)  g^{\mu\nu} \ .
\ee
It follows that the last term in the right-hand side of the above equation be interpreted as an effective  cosmological constant
\be
\Lambda_{eff} = -\frac{1}{2} ( \mathfrak{s}-1) \left(R + 16 \pi p \right)   \ ,
\ee
which for a general metric depends explicitly on the spacetime coordinates. In Einstein theory, effective cosmological constants  appear very often and, when they depend on the coordinates, in order to be able to determine their dynamics, it is necessary to take into account additional fields. In the case of the stochastic Einstein equations, we see that no additional fields are necessary. One can say that the dynamics of $\Lambda_{eff}$ is dictated by the stochastic character of the spacetime metric. 
Moreover, by using the trace equation (\ref{spur}) in the expression for the effective cosmological constant, we obtain
\be
\Lambda_{eff} = -4\pi \frac{ \mathfrak{s}-1}{1-2 \mathfrak{s}} (p-\epsilon) \ .
\ee
We see again that stiff matter ($p=\epsilon)$ is not affected by quantum fluctuations of the metric. 


\section{Stochastic FLRW cosmology}
\label{sec:flrw}
 
Let us consider the FLRW metric in the form
\be
ds^2 = dt^2 - a^2(t)\left[\frac{dr^2}{1-k r^2} + r^2 (d\theta^2 + \sin^2 \theta d\varphi^2)\right] \ .
\ee
The corresponding  stochastic field equations (\ref{see}) reduce to  
\be
( \mathfrak{s}-1)\frac{\ddot a}{a} + \frac{ \mathfrak{s}}{a^2} (\dot a ^2+k) = \frac{8\pi}{3}[ \epsilon + (1- \mathfrak{s}) p] \ ,
\label{eq1}
\ee
\be
(3 \mathfrak{s}-1)\frac{\ddot a}{a} + \frac{3 \mathfrak{s}-2}{a^2} (\dot a ^2+k) = - 8\pi   \mathfrak{s} p \ .
\label{eq2}
\ee
In the limiting case $ \mathfrak{s}=1$, we obtain the standard Friedmann equations, where Eq.(\ref{eq1}) reduces to a first-order differential equation, which is known as the Hamiltonian constraint, and Eq.(\ref{eq2}) determines the dynamics of the scale factor. On the contrary, in the case of the stochastic field equations, we have two second-order differential equations that dictate the dynamics of the expansion. However, in the stochastic case we can recover the standard structure of the Friedmann equations. Indeed, substituting the term $(\dot a ^2+k)$ of Eq.(\ref{eq1}) into Eq.(\ref{eq2}), we obtain
\be
(2 \mathfrak{s}-1) \frac{ \ddot a}{a} = -\frac{4\pi}{3} [ (3 \mathfrak{s}-2) \epsilon + (5 \mathfrak{s}-2) p] \ .
\label{deq}
\ee
Furthermore, the Hamiltonian constraint can be expressed as 
\be
\frac{ \dot a ^2 }{a^2} + \frac{k}{a^2} = \frac{4\pi}{3(2 \mathfrak{s}-1)} [(3 \mathfrak{s}-1)\epsilon + ( \mathfrak{s}-1) p ]\ .
\label{hams}
\ee
For the further analysis of these equations it is necessary to introduce {\it ad hoc} an additional equation, for instance, in the form of an equation of state 
$p=p(\epsilon)$. 

A straightforward computation of the Bianchi identities (\ref{bianchi}) with the FLRW metric shows that only one equation is not identically satisfied, namely
\be
(6 \mathfrak{s}^2-9 \mathfrak{s}+3)(\dot a^2 + k) \dot a + 2\pi  \mathfrak{s} a^3 [( \mathfrak{s}-1)\dot p + (3 \mathfrak{s}-1 ) \dot \epsilon ]
 + 4 \pi a^2 \dot a [(3 \mathfrak{s}^2 + \mathfrak{s}-1)\epsilon + (5 \mathfrak{s}^2 -  \mathfrak{s} -1) p] = 0.
\label{claw}
\ee
This equations reduces in the limiting case $\mathfrak{s}=1$  to well-known expression for the conservation law 
\be
 \dot \epsilon + 3 \frac{\dot a}{a} (\epsilon + p) = 0 \ .
\label{claweins}
\ee
The integration of the last differential equation permits us to express $\epsilon$ in terms of the scale factor $a$, when a barotropic equation of state is assumed. In the general stochastic case, however, this is not possible. Indeed, substituting the Ansatz $\epsilon = a^\alpha$ ($\alpha = const$) with a barotropic equation of
state into Eq.(\ref{claw}), one can show that there is no $\alpha$ which fulfills the resulting equation.

One of the goals of quantum gravity is to find a procedure to avoid curvature singularities which imply the break down of the classical gravity. For the model under consideration, this is equivalent to asking whether the presence of the stochastic factor $\mathfrak{s}$ in the field equations can be used to avoid the initial singularity at $t=0$ of the standard cosmological model. To investigate this question, we simply impose the condition that the solutions of the stochastic Friedmann equations (\ref{deq}) and (\ref{hams}) bounce near $t=0$.  The condition for a bouncing can be expressed 
as $\ddot a(0) > 0$, $\dot a(0) =0$ and $a(0)=a_0 = const$. Then, if we assume a barotropic equation of state $p=\omega \epsilon$ with $\epsilon(0)=\epsilon_0>0$,
the dynamic equation (\ref{deq}) implies that
\be
\frac{1}{2 \mathfrak{s}-1}[(3 \mathfrak{s}-2)+\omega ( \mathfrak{s}-2)] < 0 \ .
\ee
This inequality has two possible solutions
\be
\frac{2(1+\omega)}{3+5\omega} <  \mathfrak{s} < \frac{1}{2} \ , \qquad {\rm  i.e.} \qquad \omega \in (-\infty, -3/5) \cup (1,+\infty)\ ,
\ee
and 
\be 
 \frac{1}{2}<  \mathfrak{s} < \frac{2(1+\omega)}{3+5\omega} \ , \qquad {\rm i.e.}\qquad \omega \in (-3/5,1)\ .
\ee
In both cases, there exists a range of values for $\omega$ that satisfies the bouncing condition.

We conclude that given a stochastic factor $\mathfrak{s}$, 
there is always a  non-trivial probability that the initial singularity is not reached, and the bouncing condition
selects, by means of the barotropic factor, the type of cosmological models that represent a bouncing universe. 
For instance, if we assume that
 $ \mathfrak{s}\approx 1 $, as the result of infinitesimal quantum fluctuations, then the bouncing condition leads to $\omega \approx -1/3$. This result is physically reasonable since negative barotropic factors can be associated with repulsive gravitational effects, which are expected to be the physical cause of the bouncing.

\section{Final remarks}
\label{sec:con}

In this work, we propose a simple model to take into account in Einstein gravity the fluctuations which appear as a result of the existence of a quantum spacetime volume.  To construct our model, we assume that the quantum spacetime fluctuations can be considered in the variational principle as a stochastic factor, 
affecting only the terms which include the variation of the spacetime volume. We obtain as a result a set of stochastic classical equations which in the case of the Einstein equations with a perfect-fluid source formally correspond to multiplying the spacetime metric by the stochastic factor.  Our model can be considered as describing the behavior of classical fields on the background of a fluctuating spacetime. 

We investigate the classical stochastic classical Einstein equations in the case of the FLRW spacetime metric, and show that the corresponding Friedmann equations 
can be used to generate different cosmological solutions. We investigate the conditions under which the initial big bang singularity can be replaced by a bouncing cosmological factor. It is shown that in general there is always a non-trivial probability that the big bang is avoided. 

We conclude that it is possible to construct simple models that take into account the presence of a quantum spacetime by considering only the physical consequences of quantum fluctuations through a simple stochastic factor. In addition, our results show that considering only a stochastic factor it is possible to avoid the initial big bang singularity in a FLRW cosmology. We interpret this result as an indication that classical singularities 
can be avoided even within a stochastic model that includes quantum effects in a very simple manner.   

In our model, we assumed that quantum fluctuations lead to the appearance of a constant stochastic factor at the level of the spacetime volume, affecting the dynamics of the gravitational field. It seems possible to generalize this approach to include a stochastic function, which is probably a more realistic approach from a physical point of view. In this case, however, it will be necessary to consider the dynamics of the stochastic function at the level of the field equations. One would then expect that this type of stochastic dynamics should be related to the quantum origin of the fluctuations. We expect to investigate this problem in a future work.

\section*{Acknowledgments}
This work was partially supported by DGAPA-UNAM, Grant No. 113514, and Conacyt, Grant No. 166391. 
 VD is grateful for a grant $\Phi.0755$  in fundamental research in natural sciences by the Ministry of Education and Science of Kazakhstan.

\end{document}